\documentclass[twocolumn,showpacs]{revtex4}
\usepackage{graphicx}
\usepackage{dcolumn}
\usepackage{bm}
\usepackage{epsfig}

\input{tcilatex}

\begin{document}

\title{Breakup of $^{8}$B and the $S_{17}$ astrophysical factor revisited}
\author{L. Trache$^{1}$}
\email{l-trache@tamu.edu}
\author{F. Carstoiu$^{1,2}$}
\author{C.A. Gagliardi$^{1}$}
\author{R. E. Tribble$^{1}$}
\affiliation{$^{1}$Cyclotron Institute, Texas A\&M University, College Station, Texas
77843, USA \\
$^{2}$National Institute of Physics and Nuclear Engineering H. Hulubei,
Bucharest, Romania }
\date{\today}

\begin{abstract}
Existing experimental data for the breakup of $^{8}$B at energies from 30 to
1000 MeV/nucleon on light through heavy targets are analyzed in detail in
terms of an extended Glauber model. The predictions of the model are in
excellent agreement with independent reaction data (reaction cross sections
and parallel momentum distributions for core like fragments). Final state
interactions have been included in the Coulomb dissociation component. We
extract asymptotic normalization coefficients (ANC) from which the
astrophysical factor $S_{17}(0)$ for the key reaction for solar neutrino
production, $^{7}$Be(p,$\gamma $)$^{8}$B, can be evaluated. Glauber model
calculations using different effective interactions give consistent, though
slightly different results. The differences give a measure of the precision
one can expect from the method. The unweighted average of all ANCs extracted
leads to $S_{17}(0)=18.7\pm 1.9$ eVb. The results of this new analysis are
compared with the earlier one. They are consistent with the values from most
direct measurements and other indirect methods.
\end{abstract}

\maketitle

% 
%\keywords{Suggested keywords}%

\section{\label{sec:level1}Introduction}

The major source of the high-energy neutrinos observed by the solar neutrino
detectors is $^{8}$B, produced in the $^{7}$Be($p$,$\gamma $)$^{8}$B
reaction \cite{bahcall} at the end of the \textit{pp III} chain. The recent
results from Superkamiokande \cite{SKami} and SNO \cite{SNO} shift the
interest for a precise determination of the rate of this reaction from the
problem of the existence of the solar neutrino deficit and of the neutrino
oscillations to that of putting stringent constraints on the different
scenarios that explain them. There were many recent determinations of $%
S_{17} $, but its precise value is still controversial. In particular, there
is a discrepancy between the value found in one direct measurement and most
of the results from indirect measurements.

Recently we have proposed an indirect method to extract astrophysical $S$
factors from one-nucleon-removal (or breakup) reactions of loosely bound
nuclei at intermediate energies \cite{tra01,tra02}. It is based on the
recognition that the structure of halo nuclei is dominated by one or two
nucleons orbiting a core (see for example \cite{tanihata96,hansen01} and
references therein). Consequently, we use the fact that the breakup of halo
or loosely bound nuclei is essentially a peripheral process, and therefore
the breakup cross-sections can give information about the wave function of
the last proton at large distances from the core. More precisely, we
determine asymptotic normalization coefficients (ANCs) from a comparison of
the experimental data with calculations. Then, these ANCs are sufficient to
determine the astrophysical $S$ factors for radiative proton capture
reactions. The approach offers an alternative and complementary technique to
extracting ANCs from transfer reactions \cite{mukh01}, an alternative
particularly well adapted to rare isotope beams produced using fragmentation.

In this paper we discuss the use of existing experimental data on $^{8}$B
breakup at energies between 30 and 1000 MeV/nucleon \cite%
{nego,blank,enders,warner,cortina} to determine the astrophysical factor $%
S_{17}$. The calculations presented in \cite{tra01} on this subject were
extended and refined. First, the Coulomb part of the dissociation cross
section was modified by including the final state interaction into the
calculations. Second, new data on the breakup of $^{8}$B are analyzed \cite%
{enders,warner,cortina}. Third and most important, a new set of calculations
for the breakup of $^{8}$B were made using different effective
nucleon-nucleon (NN) interactions. Each of the new effective interactions
considered, which do not involve any new parameters, give consistent results
for all experiments, but the average ANCs found are slightly different from
one interaction to another. We interpret these differences as a measure of
the accuracy of the present (and possibly other) indirect method(s).
Finally, a brief comparison with results of direct measurements and of other
determinations of $S_{17}$ using indirect methods is made.

\section{From breakup cross sections to ANCs}

In the breakup (one-nucleon removal reactions) of loosely bound nuclei at
intermediate energies, a nucleus $B=(Ap)$, where $B$ is a bound state of the
core $A$ and the nucleon $p$, is produced by fragmentation from a primary
beam, separated and then used to bombard a secondary target. In inclusive
measurements, the core $A$ is detected, measuring its parallel and
transverse momenta and eventually the gamma-rays emitted from its
deexcitation. Spectroscopic information can be extracted from these
experiments, such as the orbital momentum of the relative motion of the
nucleon and the contribution of different orbitals (from the momentum
distributions) and the contribution of different core states (from the
coincidences with gamma-rays). Typically the experimental results are
compared with calculations using Glauber models. The integrated cross
sections have been used to extract absolute spectroscopic factors \cite%
{hansen01} or the ANC \cite{tra01}. We have shown that the latter approach
has the advantage that it is independent of the geometry of the proton
binding potential, an important feature for exotic nuclei for which the
geometry of the mean field is not necessarily well known. The ANC $%
C_{Ap}^{B} $ for the nuclear system $A+p\leftrightarrow B$ specifies the
amplitude of the tail of the overlap function of the bound state $B$ in the
two-body channel $(A\,p)$ \cite{mukh01}. This ANC is enough to determine the
direct (non-resonant) contribution to the astrophysical $S$ factor for the
radiative proton capture reaction $A(p,\gamma )B$ which is a highly
peripheral process due to the Coulomb barrier and the low energies in the
entrance channel. Using this strategy we described the breakup of $^{8}$B in
terms of an extended Glauber model. The $^{8}$B projectile (made of a proton
and the $^{7}$Be core) is moving on a straight line trajectory and each part
is interacting independently with the target. The breakup cross sections
depend on the proton-target and core-target interactions and on the relative 
$p$-core motion.

The wave function of the ground state of $^{8}$B is a mixture of $1p_{3/2}$
and $1p_{1/2}$ orbitals, around a $^{7}$Be core: 
\begin{eqnarray}
|^{8}B(g.s.) &>&=A_{p_{3/2}}\left[ ^{7}Be(3/2^{-})\otimes p_{_{3/2}})\right]
_{2^{+}}+  \nonumber \\
&&A_{p_{1/2}}\left[ ^{7}Be(3/2^{-})\otimes p_{1/2})\right] _{2^{+}}+ 
\nonumber \\
&&A_{e}\left[ ^{7}Be^{\ast }(1/2^{-})\otimes p_{3/2})\right] _{2^{+}}+...
\label{B8-str}
\end{eqnarray}%
where $A_{i}$ are the spectroscopic amplitudes of the various components.
The first two terms represent the proton in the $1p_{3/2}$ and $1p_{1/2}$
orbitals, respectively, coupled to the ground state of $^{7}$Be. The third
term corresponds to the proton being coupled to the first excited state of
the $^{7}$Be core, at E$^{\ast }$=0.429 MeV. Basic shell model arguments
suggest that the $1p_{3/2}$ term dominates, and only a small $1p_{1/2}$
admixture exists. Recently, in the study of its mirror nucleus $^{8}$Li, we
disentangled for the first time these two contributions and found their
ratio to be $%
A_{p_{1/2}}^{2}/A_{p_{3/2}}^{2}=C_{p_{1/2}}^{2}/C_{p_{3/2}}^{2}=0.13(2)$\ 
\cite{tra03}. Only these two terms contribute in the radiative capture
process. However, all three terms contribute in the breakup process, with
the third one identified in $^{8}$B breakup through coincidences with
gamma-rays \cite{cortina}. It does not contribute in the radiative capture,
but its contribution has to be evaluated and subtracted from all the other
inclusive breakup data. From the breakup cross section to the excited state
in $^{7}$Be, $\sigma (exc.)=12(3)$ mb and $\sigma (tot)=94(9)$ mb measured
at 936 MeV/nucleon, we found $C_{e}^{2}/C_{tot}^{2}=0.16(4),$ a value
consistent with that found in the original analysis in \cite{cortina} and
which, subsequently, was used to correct for the contribution of core
excitation in all other breakup data analyzed here. These two findings
together establish the wave function of the ground state of $^{8}$B, up to
an overall multiplicative factor.

The calculated one-proton removal cross sections and the momentum
distributions are given by the incoherent superposition of the single
particle contributions from the different parts of the wave function
weighted by the respective spectroscopic factors \cite{tostevin01} 
\begin{equation}
\sigma _{-1p}=\sum S(c,nlj)\sigma _{sp}(nlj).  \label{cs-tot}
\end{equation}%
In inclusive measurements, such as those analyzed here, the proton is not
detected, therefore the calculated cross sections $\sigma _{sp}(nlj)$
contain a stripping term (the loosely bound proton is absorbed by the target
and the $^{7}$Be core is scattered and detected), a diffraction dissociation
term (the proton is scattered away by the target, the $^{7}$Be core is
scattered by the target and is detected) and a Coulomb dissociation term 
\cite{hencken} 
\begin{equation}
\sigma _{sp}=\int_{0}^{\infty }2\pi bdb(P_{str}(b)+P_{diff}(b))+\sigma
_{Coul}  \label{sig-sp}
\end{equation}

These terms were calculated using the extended Glauber model detailed
elsewhere \cite{sauvan,carstoiu04}. S-matrix elements have been calculated
in the eikonal approximation up to the second order \cite{wallace} to assure
convergence. This convergence was checked with calculations for other
quantities, for example proton-target reaction cross sections as a function
of energy, and compared with data available from literature \cite{NNDC}.

\section{Results using different NN interactions}

\begin{figure}[tbh]
\begin{minipage}{80mm}
    \mbox{\epsfxsize=7.9cm\epsfbox{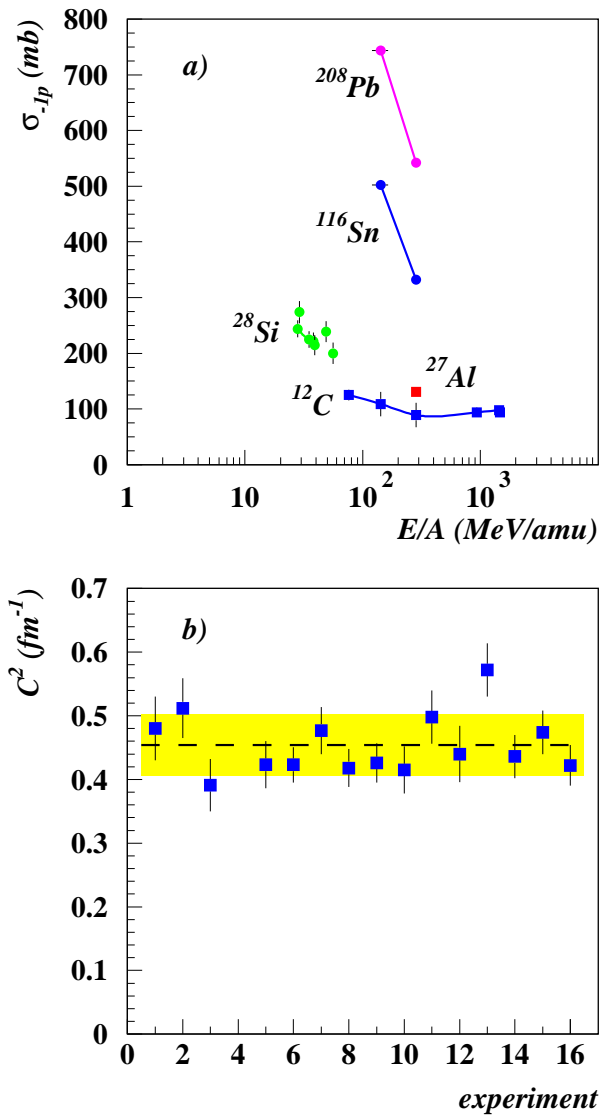}}
\caption{a) The cross sections determined from the breakup of $^{8}$B at
30-1000 MeV/u on C, Al, Sn and Pb targets at various energies [9-13] used
in this study. b)The ANCs determined from the breakup of $^{8}$B using the JLM effective interaction. The error bars of the individual points contain the experimental and theoretical uncertainties. The dashed line shows the average and the hatched area is the standard deviation.  }
\label{fig1}
\end{minipage}
\end{figure}

In calculations we assume a structure of the projectile given by Eq. \ref%
{B8-str}, with the spectroscopic factors, or the ANCs, to be determined from
the comparison of the measured cross sections (from which the contribution
of the $^{7}$Be core excitation was removed as described above) with those
calculated as an incoherent superposition of single particle cross sections%
\begin{equation}
\sigma _{-1p}=(S_{p_{3/2}}+S_{p_{1/2}})\sigma
_{sp}=(C_{p_{3/2}}^{2}+C_{p_{1/2}}^{2})\sigma _{sp}/b_{p}^{2},  \label{SC2}
\end{equation}%
where $b_{nlj}$ are the asymptotic normalization coefficients of the
normalized single particle radial wave functions $\varphi _{nlj}(r)$\
calculated in a spherical Woods-Saxon potential of a given geometry and with
the depth adjusted to reproduce the experimental proton binding energy of $%
^{8}$B, $S_{p}=0.137$ MeV. They are essentially equal for the $1p_{3/2}$ and 
$1p_{1/2}$ orbitals ($b_{p}$), as are the single particle breakup cross
sections $\sigma _{sp}$. The sum of the spectroscopic factors or the sum of
the asymptotic normalization coefficients $%
C_{tot}^{2}=C_{p_{3/2}}^{2}+C_{p_{1/2}}^{2}$ can thus be extracted by
comparing the experimental one-proton removal cross sections with the
calculations. The $^{8}$B ANC, $C_{tot}^{2}$, is extracted from existing
breakup data at energies between 30-1000 MeV/nucleon and on different
targets ranging from C to Pb \cite{nego,blank,enders,warner,cortina}. Figure %
\ref{fig1} a) shows the one-proton removal cross sections for various
targets and incident energies. One can notice the large range of cross
sections and the variation with the energy for different targets. 

Two approaches were used to evaluate the $S$-matrices needed in the
calculations. The first is a potential approach. To obtain the folded
potentials for the proton-target and core-target interactions we used the
effective nucleon-nucleon interaction of Jeukenne, Lejeune and Mahaux (JLM) 
\cite{JLM} and Hartree-Fock-Bogolyubov densities carefully adjusted to
correctly reproduce the experimental binding energy of each nucleus. In an
extensive study of the elastic scattering of loosely bound \textit{p}-shell
nuclei around 10 MeV/u \cite{tra00}, we found that renormalized double
folded potentials with this effective interaction provide a good description
of the data. We found there that a large renormalization is needed for the
real part of the potential, but no renormalization is needed for the
imaginary part of the potential. In the present calculations we assume that
no renormalization of the imaginary part is needed at all energies. We used
the JLM interaction for energies below 285 MeV/nucleon only.

\begin{figure}[tbh]
\begin{minipage}{80mm}
    \mbox{\epsfxsize=8.9cm\epsfbox{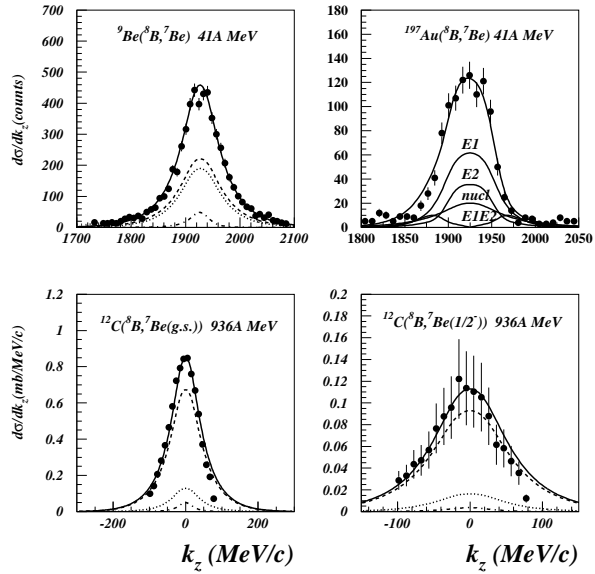}}
\caption{The parallel momentum distributions determined from the breakup 
of $^{8}$B on Be and Au targets at 41 MeV/u \protect\cite{kelley} 
and on C at 936 MeV/u \protect\cite{cortina} for the both g.s. and core excitation components. Final state interaction is included in the Coulomb calculations. The total (full lines) and the components are shown: stripping (dashed) and diffraction (dotted), Coulomb (dash-dotted), or as labelled on each curve. }
\label{fig2}
\end{minipage}
\end{figure}

\begin{figure}[tbh]
\begin{minipage}{80mm}
    \mbox{\epsfxsize=7.9cm\epsfbox{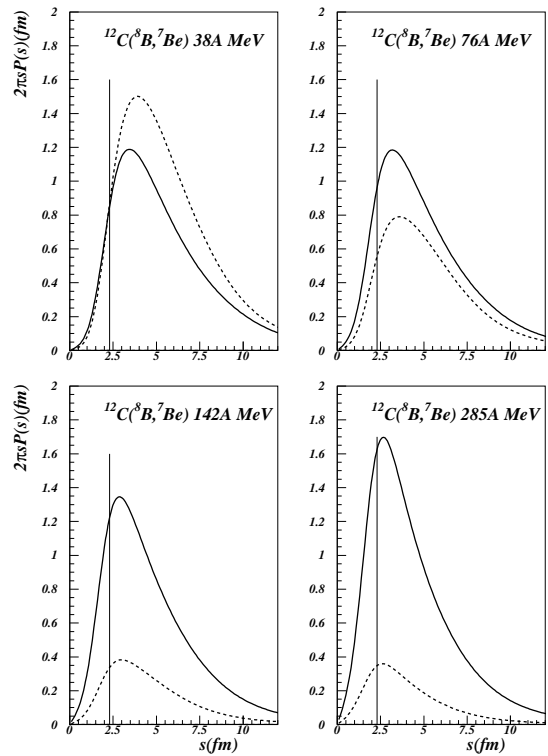}}
\caption{The breakup probability profiles as a function of the impact
paramenter \textit{s} for the breakup of $^{8}$B on C targets at four
different energies. The stripping (full lines) and the diffraction 
dissociation (dashed lines) components are shown. The vertical line shows the position of the $^7$Be core {\it rms} radius.}
\label{fig3}
\end{minipage}
\end{figure}

Before comparing the experimental and calculated integrated cross sections,
we checked that we can reproduce all other available experimental
observables with our model. This was crucial before proceeding with the
calculations. In Figure \ref{fig2} we show that parallel momentum
distributions measured at 41 MeV/nucleon on one low Z (Be) and one high Z
(Au) target \cite{kelley} and on the $^{12}$C target at 936 MeV/nucleon
(calculated with appropriate technique for high energy, as discussed below)
for both the ground state and excited state components are well reproduced.
Similarly, the transverse momentum distributions are well reproduced. The
model also reproduces well the relative fraction of stripping/diffraction
dissociation disentangled first by Negoita et al. \cite{nego} on Si targets
at 28-38 MeV/nucleon (as can be seen in Fig. 6 of that reference) and more
recently by Enders et al. on C at 76 MeV/nucleon \cite{enders}. For the
latter the calculations give $\sigma _{str}=80$ mb, $\sigma _{diff+C}=50$
mb, to compare with the experimental results $\sigma _{str}=93(16)$ mb, $%
\sigma _{diff+C}=37(13)$ mb. In Figure \ref{fig3} we show, for the case of
the breakup of $^{8}$B on C targets that the reaction is essentially
peripheral. The stripping and the nuclear diffraction dissociation
probabilities as a function of the proton impact parameter, $s,$ are
calculated at four energies. While these probabilities are peaked outside
the radius of the $^{7}$Be core (vertical line) in all cases, it is clear
that the interior contributes and should be carefully considered. The figure
also shows the variation with energy of the relative importance of the two
nuclear mechanisms: the diffraction dissociation (dashed line) is dominant
at lower energies and its role decreases with increasing energy where
stripping (full line) becomes dominant. The comparison of the results of the
present calculations with the results of the simpler black disk model shows
that the interior plays the crucial role in describing correctly the wings
of the parallel momentum distributions (see Fig. 3 of Ref. \cite{tra01}). An
analysis like the one presented in Fig. 2 of Ref. \cite{tra01} shows that
there is an energy window E/A=25-150 MeV/nucleon for which the breakup of $%
^{8}$B is mostly peripheral even on the lightest targets. For the heavier
targets this is always the case, due to the dominance of the Coulomb
component.

The data considered were taken on C targets at 76 \cite{enders}, 142, 285 
\cite{blank} and 936 A MeV \cite{cortina} (exps. no. 1-4, in order), on Al
at 285 A MeV (\cite{blank}, exp. \#5), on Si at 28, 35, 38 (\cite{nego},
exps. \#6,8,9), 29, 39, 49 and 56 A MeV (\cite{warner}, exp. \#7,10-12), on
Sn at 142 and 285 A MeV (\cite{blank}, exps. \#13,14) and on Pb targets at
142 and 285 A MeV (\cite{blank}, exps. \#15,16). 

From the analysis with the JLM interaction of all experiments up to 285 A
MeV we find ANCs consistent with a constant value (Fig. \ref{fig1} b) with
an average $C_{tot}^{2}(JLM)=0.454\pm 0.048$ fm$^{-1}$. Compared with Ref. 
\cite{tra01}, we include the newer measurements by \cite{enders,warner}.
Another distinction is that we have included the final state interaction in
the calculation of the Coulomb dissociation component of the one-proton
removal cross section. E1 and E2 amplitudes have been included as in the
earlier calculation, except that distorted waves, not plane waves, were
taken in the $p+^{7}$Be\ final channel for the calculation of the matrix
elements. The distorted waves were calculated numerically in the same
potential that was used to bind the proton $p$ around the $^{7}$Be core in
the ground state of $^{8}$B. Differences occur between the calculated
amplitudes with the two approaches especially for low relative momenta, but
their influence on the final integrated result is relatively small due to
the extra $q^{2}$ factor that weights\ their contribution to the integrated
cross section.\ However, the inclusion of distorted waves increases the
asymmetry in the parallel momentum distribution due to an increased E1-E2
interference effect as can be seen in the upper right panel in Fig \ref{fig2}%
. It has been suggested \cite{davids01} that asymmetries observed in the
fragment parallel momentum distributions in the Coulomb dissociation of $^{8}
$B on heavy targets could be reproduced with an overall renormalization of
1.22 and of 0.7 for the E2 matrix elements calculated in first order
perturbation theory. We have, therefore, performed calculations using bare
amplitudes resulting from perturbation theory \cite{baur}, as well as
renormalized E2 and E1 amplitudes.\ No significant differences were found in
the extracted ANCs with these two versions, and the values reported here are
those obtained without any renormalization. The Coulomb term in the breakup
cross section is particularly important for heavy targets where it becomes
dominant. The value found above for the ANC is in very good agreement with
that determined before using the peripheral proton transfer reactions $^{10}$%
B($^{7}$Be,$^{8}$B)$^{9}$Be and $^{14}$N($^{7}$Be,$^{8}$B)$^{13}$C at 12
MeV/nucleon \cite{azhari} $C_{tot}^{2}(p)=0.449\pm 0.045$ fm$^{-1}$ and with
that obtained from the study of the mirror neutron transfer reaction ($^{7}$%
Li,$^{8}$Li) $C_{tot}^{2}(n)=0.455\pm 0.047$ fm$^{-1}$ \cite{tra03}. They
agree very well, in spite of the differences in the energy ranges and in the
reaction mechanisms involved. The ANC extracted with JLM leads to the
astrophysical factor $S_{17}(0)=17.5\pm 1.8$ eV$\cdot $ b for the key
reaction for solar neutrino production$^{7}$Be($p$,$\gamma $)$^{8}$B. 

In a second approach, the Glauber model in the optical limit\ \cite%
{al-khalili} was used. The breakup process is treated as multiple elementary
interactions between the partners' nucleons. The total NN cross sections and
the scattering amplitudes are taken from literature. Calculations were done
for all the experiments in the energy range 50-1000 MeV/nucleon using a
constant ("standard") finite range of 1.5 fm, as well as specific ranges in
each NN channel as suggested by Ray \cite{ray}. No new parameters were
adjusted. For details on the procedure see \cite{tra02}. 
\begin{figure}[tbh]
\begin{minipage}{80mm}
   \mbox{\epsfxsize=7.cm\epsfbox{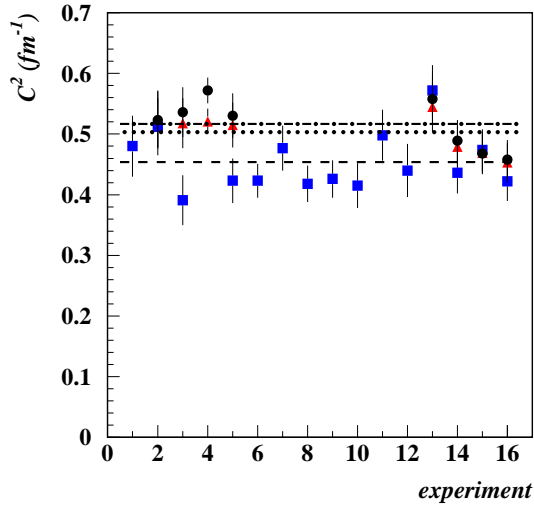}}
\caption{The ANCs determined from the breakup of $^{8}$B at 30-1000 MeV/u on
C, Al, Sn and Pb targets, using three NN effective interactions: JLM 
(squares), "standard" (circles) and "Ray" (triangles). 
See text for details. The dashed, dotted and dash-dotted lines are the average of the JLM, "standard" and "Ray" values, respectively.}
\label{fig5}
\end{minipage}
\end{figure}
For all the effective NN interactions we checked that they correctly
describe complementary data, like proton-target and $^{7}$Be-target elastic
and total reaction cross sections, where available. Understandably,
calculations with zero-range and with 2.5 fm range give too small or too
large cross sections, respectively, and were not retained. Data at energies
higher than 50 A MeV were selected. We did not include the measurements of
Ref. \cite{smedberg} at 1440 MeV/nucleon and of Ref. \cite{schwab} at 1471
MeV/nucleon (highest energy points in Fig. \ref{fig1}a), because at those
very large energies the breakup is no longer peripheral and the extraction
of an ANC may not be the most appropriate. However, the results obtained
from the analysis of these two cases are fully consistent with the others. \ 

For each of the two NN interactions we find that all experiments give
consistent ANCs (Fig. \ref{fig5}), but the average values obtained are
slightly different: $C_{tot}^{2}($"\textit{standard}"$)=0.503\pm 0.032$ fm$%
^{-1}$ and $C_{tot}^{2}($\textit{"Ray"}$)=0.517\pm 0.041$ fm$^{-1}$. These
differ by 11\% and 13\% respectively from the JLM value. We find no argument
to determine which value is best. If we take the unweighted average of all
31 determinations we find an ANC $C_{tot}^{2}($\textit{ave}$)=0.483\pm 0.050$
fm$^{-1}$ that leads to $S_{17}(0)=18.7\pm 1.9$ eV$\cdot $ b. The
uncertainties quoted are only the standard deviation of the individual
values around the averages, with no experimental errors included. The
experimental data considered here were taken in various laboratories, at
different energies, with varying methods and the calculations also used
different techniques. Therefore, we believe that the results form a
statistical ensemble with many and randomly occurring error sources, for
which the average and the standard deviation around the average give a
reasonable description of the ANC and its error.

In Ref. \cite{brown02} the authors study the same data of $^{8}$B breakup on
the C target and find a larger value for the ANC than the one we published
previously in \cite{tra01}. They use a different strategy for the
calculations where they assume a wave function for the $^{8}$B g.s. from
nuclear structure calculations and a geometry of the proton binding
potential that they do not question. Then, the comparison with the
experiment gives them a quenching factor $R_{s}$ of unexplained origin in
that paper (but of great significance if its connection with short range
correlations inside nuclei is confirmed). On the other hand they compare
their result for one single target with the full average from our
calculations. A direct comparison with the individual ANCs or with the
average of our results for the breakup on the C target only (available in
Table I of our Ref. \cite{tra01}) would have led to agreement. Later \cite%
{enders} they find full agreement with us \cite{tra02} for the breakup of $%
^{9}$C where we use essentially the same techniques. %Actually,
%when we recalculate $C_{tot}^{2}$ with the explicit formula given in Eq. 3
%of \cite{enders} and their results from table II of Ref. \cite{brown02} we
%find $C_{tot}^{2}=0.51,$ and not $0.57$ as they report in the text and in
%agreement within the error bars with our result \cite{tra01}. 
Also, our examination of different theoretical reaction models above
indicates that a quenching factor $R_{s}=0.88$ may not be precise enough to
consider it different from unity. A recent study of 23 cases of one-neutron
removal cross sections at similar energies \cite{sauvan} found no quenching $%
R_{ave}=0.98\pm 0.16$.

\section{\protect\bigskip Conclusions}

In conclusion, we show that the breakup of $^{8}$B at intermediate energies
can be used to obtain the $S_{17}$ astrophysical factor at stellar energies.
Very difficult direct measurements are complemented by reactions using
secondary beams of exotic nuclei obtained from fragmentation and seeking the
relevant ANCs, rather than a complete knowledge of the ground state wave
function of $^{8}$B. In addition, the indirect ANC method is subject to
different systematic errors than direct measurements.

There were many recent determinations of this key astrophysical factor $%
S_{17}$, but its precise value is still controversial. Our result is in
agreement with those from all indirect methods and with most of the direct
determinations (see the discussions in \cite%
{schumann,davids03,hammache,strieder}), but one which stands out in its
claim of a larger value and very small error \cite{junghans}. The value
obtained as an average of all ANCs found in the present study $%
S_{17}(0)=18.7\pm 1.9$ eV$\cdot $ b, is virtually equal with the most
probable values extracted in Ref. \cite{schumann} $S_{17}(0)=18.6\pm 1.2$
(stat) $\pm 1.0$ (theor) eV$\cdot $ b and in Ref. \cite{davids03} $%
S_{17}(0)=18.6\pm 0.4$ (stat) $\pm 1.1$ (theor) eV$\cdot $ b from
statistical analyses of all mutually consistent results, including the
reanalysis of data from direct measurements \cite{baby} with a different
extrapolation at low energies. Our results from the use of different NN
interactions reminds us of the fact that the precision of indirect methods
depends not only on the precision of the experiments but also on the
accuracy of the calculations. These findings may give a measure of the
present status for break-up reactions, indicating that accuracies to +/-10\%
can be obtained. 
%However, we cannot find support for a value as large as 22 eVb.

\begin{acknowledgments}
This work was supported in part by the U. S. Department of Energy under
Grant No. DE-FG03-93ER40773, by the Romanian Ministry for Research and
Education under contract no 555/2000, and by the Robert A. Welch Foundation.
\end{acknowledgments}


\begin{thebibliography}{99}
\bibitem{bahcall} J. N. Bahcall, M. H. Pinsoneault and S. Basu, Astrophys.
J. \textbf{555} (2001) 990.

\bibitem{SKami} SuperKamiokande collaboration, S. Fukuda \textit{et al}.,
Phys. Rev. Lett. \textbf{86}, 5651 (2001).

\bibitem{SNO} SNO\ collaboration, S. N. Ahmed \textit{et al}., Phys Rev.
Lett. \textbf{87}, 071301 (2003).

\bibitem{tra01} L. Trache, F. Carstoiu, C. A. Gagliardi and R.E. Tribble,
Phys. Rev. Lett. \textbf{87, }271102 (2001).

\bibitem{tra02} L. Trache, F. Carstoiu, M. A. Mukhamedzhanov and R.E.
Tribble, Phys. Rev. C \textbf{66,} 035801 (2002).

\bibitem{tanihata96} I. Tanihata, J. Phys. G:\ Nucl. Part. Phys. \textbf{22,}
157 (1996).

\bibitem{hansen01} P. G. Hansen and B. M. Sherrill, Nucl. Phys. \textbf{A693,%
} 133 (2001).

\bibitem{mukh01} A. M. Mukhamedzhanov, C. A. Gagliardi and R. E. Tribble,
Phys. Rev. C \textbf{63,} 024612 (2001).

\bibitem{nego} F. Negoita \textit{et al.}, Phys. Rev. C \textbf{54,} 1787
(1996).

\bibitem{blank} B. Blank \textit{et al.}, Nucl. Phys. \textbf{A624,} 242
(1997).

\bibitem{enders} J. Enders \textit{et al.}, Phys. Rev. C \textbf{67,} 064301
(2003).

\bibitem{warner} R. E. Warner \textit{et al.}, Bull. Am. Phys. Soc. \textbf{%
47}, 29 (2002) and private communication.

\bibitem{cortina} D. Cortina-Gil \textit{et al.}, Nucl. Phys. \textbf{A720, }%
3 (2003).

\bibitem{tra03} L. Trache \textit{et al.}, Phys. Rev. C \textbf{67},
062801(R) (2003).

\bibitem{tostevin01} J. A. Tostevin, Nucl. Phys. \textbf{A682}, 320c (2001).

\bibitem{hencken} K. Hencken, G. Bertsch and H. Esbensen, Phys. Rev. C 
\textbf{54}, 3043 (1996).

\bibitem{sauvan} E. Sauvan \textit{et al.}, nucl-ex/0307019, to be published
in Phys Rev. C.

\bibitem{carstoiu04} F. Carstoiu \textit{et al.}, to be published.

\bibitem{wallace} S. J. Wallace, Phys. Rev. C 8, 2043 (1973).

\bibitem{NNDC} National Nuclear Data Center, Brookhaven National Laboratory
Online Data Service.

\bibitem{JLM} J. P. Jeukenne, A. Lejeune and C. Mahaux, Phys. Rev. C \textbf{%
16, 80} (1977).

\bibitem{tra00} L. Trache \textit{et al.} Phys. Rev. C \textbf{61}, 024612
(2000).

\bibitem{kelley} J. H. Kelley \textit{et al.}, Phys. Rev. Lett. \textbf{77},
5020 (1996).

\bibitem{davids01} B. Davids \textit{et al.}, Phys. Rev. C \textbf{63},
065806 (2001).

\bibitem{baur} C. Bertulani and G. Baur, Nucl. Phys. \textbf{A480}, 615
(1988).

\bibitem{azhari} A. Azhari \textit{et al.}, Phys. Rev. Lett. \textbf{82},
3960 (1999); Phys. Rev. C \textbf{60,} 055803 (1999).

\bibitem{al-khalili} J. S. Al-Khalili and J. A. Tostevin, Phys. Rev. Lett. 
\textbf{76}, 3903 (1996).

\bibitem{smedberg} M. H. Smedberg \textit{et al.}, Phys. Lett. B \textbf{452}%
, 1 (1999).

\bibitem{schwab} W. Schwab \textit{et al.}, Z. Phys. A \textbf{350}, 283
(1995).

\bibitem{ray} L. Ray, Phys. Rev. C \textbf{20,} 1857 (1979).

\bibitem{brown02} B.A. Brown, P. G. Hansen, B. M. Sherrill and J. A.
Tostevin, Phys. Rev. C \textbf{65}, 061601(R) (2002).

\bibitem{schumann} F. Schumann \textit{et al.}, Phys. Rev. Lett. \textbf{90,}
232501 (2003) and references therein.

\bibitem{davids03} B. Davids and S. Typel, Phys. Rev. C \textbf{68,} 045802
(2003).

\bibitem{hammache} F. Hammache \textit{et al.}, Phys. Rev. Lett. \textbf{86,}
3985 (2001).

\bibitem{strieder} F. Strieder \textit{et al}., Nucl. Phys. \textbf{A696},
219 (2001).

\bibitem{junghans} A. R. Junghans \textit{et al.}, Phys. Rev. Lett. \textbf{%
88,} 041101 (2002).

\bibitem{baby} L. T. Baby \textit{et al.}, Phys. Rev. C \textbf{67}, 065805
(2003).
\end{thebibliography}
\end{document}